\documentclass[english,twocolumn,reprint,prapplied,nofootinbib,longbibliography,superscriptaddress]{revtex4-2}
\usepackage[T1]{fontenc}
\usepackage[latin9]{inputenc}
\usepackage{lmodern}
\usepackage{amsmath}
\usepackage{amssymb}
\usepackage{graphicx}
\usepackage[dvipsnames]{xcolor}
\usepackage[colorlinks, breaklinks, linkcolor=OrangeRed,
citecolor=RoyalBlue, urlcolor=NavyBlue]{hyperref}
\usepackage[final]{changes}

\usepackage{babel}

\begin{document}
\title{Gain compression in Josephson Traveling-Wave
Parametric Amplifiers}

    \author{Gwenael Le Gal} 
	\affiliation{Univ. Grenoble Alpes, CNRS, Grenoble INP, Institut N\'eel, 38000 Grenoble, France}
    \author{Guilliam Butseraen} 
    \thanks{Now at Silent Waves.}
	\affiliation{Univ. Grenoble Alpes, CNRS, Grenoble INP, Institut N\'eel, 38000 Grenoble, France}
    \author{Arpit Ranadive}
    \thanks{Now at Google.}
	\affiliation{Univ. Grenoble Alpes, CNRS, Grenoble INP, Institut N\'eel, 38000 Grenoble, France}
    \author{Giulio Cappelli} 
	\affiliation{Univ. Grenoble Alpes, CNRS, Grenoble INP, Institut N\'eel, 38000 Grenoble, France}
    \author{Bekim Fazliji} 
	\affiliation{Univ. Grenoble Alpes, CNRS, Grenoble INP, Institut N\'eel, 38000 Grenoble, France}
	\affiliation{Silent Waves, 38000 Grenoble, France}
    \author{Edgar Bonet} 
	\affiliation{Univ. Grenoble Alpes, CNRS, Grenoble INP, Institut N\'eel, 38000 Grenoble, France}
    \author{Eric Eyraud} 
	\affiliation{Univ. Grenoble Alpes, CNRS, Grenoble INP, Institut N\'eel, 38000 Grenoble, France}
    \author{Luca Planat}
    \affiliation{Silent Waves, 38000 Grenoble, France}
    \author{Nicolas Roch} 
    \thanks{Corresponding author: nicolas.roch@neel.cnrs.fr}
	\affiliation{Univ. Grenoble Alpes, CNRS, Grenoble INP, Institut N\'eel, 38000 Grenoble, France}
    \affiliation{Silent Waves, 38000 Grenoble, France}

\begin{abstract}
Superconducting traveling-wave parametric amplifiers (TWPAs) are increasingly used in various applications, including quantum computing, quantum sensing, and dark matter detection. However, one important characteristic of these amplifiers, gain compression, has not received much attention. As a result, there is a lack of comprehensive experimental exploration of this phenomenon in the existing literature. In this study, we present an experimental investigation of gain compression in a Josephson traveling-wave parametric amplifier based on a four-wave mixing process. We have implemented a novel setup to monitor the complex transmission of both the pump and signal tones, which allows us to simultaneously track pump depletion and signal amplification as functions of signal power and frequency across the entire bandwidth of the device. Our findings indicate that, while pump depletion occurs during gain compression, it is not the only mechanism involved in the saturation of a TWPA. Power-induced phase-matching processes also take place within the device. This study provides valuable insights for optimizing TWPAs for applications that require high total input power, such as multiplexed qubit readout or broadband photon emission.
\end{abstract}
\maketitle

\section{Introduction}

Superconducting traveling-wave parametric amplifiers can be constructed using high-kinetic inductance materials (K-TWPA) \citep{ho_eom_wideband_2012} or Josephson junction arrays (J-TWPA)~\citep{esposito_perspective_2021}. These amplifiers provide bandwidths extending to several GHz, achieve gains of $\sim20$ dB, and exhibit noise characteristics approaching the quantum limit. Additionally, TWPAs can be located directly at the coldest stage of dilution refrigerators because of their limited power dissipation. This placement minimizes the losses between the device under test and the TWPA, thus improving the signal-to-noise ratio (SNR) of the detection~\citep{clerk_introduction_2010}. Collectively, these benefits have established TWPAs as essential amplifiers in circuit quantum electrodynamics (cQED) experiments, facilitating high-fidelity readout of superconducting~\citep{macklin_nearquantum-limited_2015,krinner_realizing_2022,ardati_using_2024,heinsoo_rapid_2018,krinner_engineering_2019, ranzani_kinetic_2018, gaikwad_entanglement_2024} and semiconducting qubits~\citep{de_jong_rapid_2021, elhomsy_broadband_2023}, and significantly increasing the SNR in quantum optics experiments~\citep{fraudet_direct_2025, osullivan_deterministic_2024}. 

A fundamental attribute of any amplifier is its saturation power, defined as the input signal power at which a deviation from linear gain emerges. The 1-dB compression point $P_\text{1dB}$ serves as a prevalent metric to assess the saturation of the amplifier. This point is the input signal power at which the amplifier's gain experiences a 1-dB reduction. A high $P_\text{1dB}$ is particularly advantageous for applications that require substantial total input power, such as in the multiplexed readout of superconducting qubits \citep{krinner_realizing_2022,remm_intermodulation_2023} or spin qubits \citep{elhomsy_broadband_2023}. Resonant Josephson parametric amplifiers (JPAs) \citep{aumentado_superconducting_2020}, representing another class of near-quantum-limited amplifiers, generally exhibit 1-dB compression points below $-120$ dBm \citep{liu_josephson_2017,planat_understanding_2019}. However, enhanced JPA designs can achieve saturation powers that reach $-90$ dBm \citep{naaman_high_2017,white_readout_2023, kaufman_simple_2024}. This advancement is attributed to the diluted non-linearity provided by arrays of Josephson junctions, or composite elements \citep{eichler_controlling_2014, planat_understanding_2019, frattini_optimizing_2018, sivak_josephson_2020, white_readout_2023, kaufman_simple_2024}, as well as the attenuation of higher-order nonlinear effects of the Josephson potential \citep{kochetov_higher-order_2015, boutin_effect_2017,liu_optimizing_2020}.

The 1-dB compression point $P_\text{1dB}$ ranges from $-110$ to $-95$ dBm for J-TWPA \citep{esposito_perspective_2021,macklin_nearquantum-limited_2015,planat_photonic-crystal_2020,ranadive_kerr_2022}, while K-TWPA can reach $-50$ dBm \citep{malnou_three-wave_2021}. The difference between JPAs and TWPAs can be understood by examining the ratio of non-linearity $ K $ to bandwidth $\kappa$~\citep{eichler_controlling_2014, planat_understanding_2019}. Most JPAs typically consist of around one hundred nonlinear elements, whereas TWPAs are usually made up of thousands of such components---Josephson junctions in the case of J-TWPA and squares for K-TWPA. This design effectively distributes the non-linearity across all the elements, thereby reducing the value of $ K $. Additionally, TWPAs offer a larger bandwidth compared to JPAs. Consequently, the ratio  $K/\kappa$ is more favorable for TWPAs when it comes to maximizing $P_\text{1dB}$. Despite the first demonstrations of TWPAs occurring over a decade ago \citep{ho_eom_wideband_2012, macklin_nearquantum-limited_2015}, a comprehensive experimental exploration of their saturation remains absent, to the best of our knowledge. Theoretical analyses have been focusing solely on the effect of pump depletion---a notable reduction in the pump tone's amplitude as energy is transferred to the signal and idler modes, to study saturation in TWPAs \citep{yaakobi_parametric_2013,obrien_resonant_2014,kow_self_2022}. Another potential constraint on TWPA power handling is cross-phase modulation prompted by high-power signals. Analogous to observations in JPA, where a strong signal can cause deviation from optimal biasing conditions, a strong signal might impair the essential phase-matching of a TWPA. Further propagative effects have to be considered as well. The final potentially detrimental effect is the generation of spurious tones by the device, including harmonics \citep{levochkina_investigating_2024}, sidebands, or intermodulation products \citep{remm_intermodulation_2023,walker_handbook_2012}. Recent research on the generation of intermodulation products in a J-TWPA \citep{remm_intermodulation_2023} has demonstrated that these unwanted tones can be harmful to multiplexed qubit readout and that their generation is intrinsically connected to the saturation properties of the amplifier, similar to what is observed in non-superconducting voltage amplifiers \citep{walker_handbook_2012}.

In this article, we conduct a comprehensive experimental study on gain compression in a J-TWPA. Although the 1-dB compression point of TWPAs is usually measured and discussed at peculiar frequencies in the amplification band \citep{esposito_perspective_2021}, we systematically measure the 1-dB compression point ($P_{\text{1dB}}$) throughout the amplifier bandwidth, anticipating significant variations due to frequency-dependent gain, which may also be affected by device imperfections and the microwave environment \citep{nilsson_peripheral_2024}. A novel measurement approach is introduced, which uses two vector network analyzers (VNA) to simultaneously measure the complex transmission of pump and signal tones. The article is structured as follows. Section \ref{sec:TWPA_experiment} briefly explains the operating principle of the device under study, a TWPA constructed with Superconducting Nonlinear Asymmetric Inductive Elements (SNAIL) operating in the reversed-Kerr regime \citep{ranadive_kerr_2022}, along with the experimental setup. Section \ref{sec:Modeling_compression} discusses the theoretical framework used to model the data using the well-known coupled-mode equations \citep{yaakobi_parametric_2013,obrien_resonant_2014,planat_resonant_2020,ranadive_kerr_2022}, comparing the results for the gain and pump transmission profiles. Section \ref{sec:Understanding_1dB} delves deeper into 1-dB gain compression and its correlation with pump depletion, but also other effects such as power-induced phase-mismatch. Lastly, in Section \ref{sec:Conclusion-and-discussions}, the broader implications of the results are discussed, along with strategies to address the current low-saturation power limitations of Josephson TWPAs.

\section{Traveling-wave parametric amplification and description of the experiment}

\label{sec:TWPA_experiment}

This study examines the saturation characteristics of a J-TWPA operating within the four-wave mixing (4WM) regime. In this context, effective amplification is contingent on two conditions: energy conservation $2\omega_{p}=\omega_{s}+\omega_{i}$ and momentum conservation $2\tilde{k}_{p}=\tilde{k}_{s}+\tilde{k}_{i}$ among the pump \textit{p}, signal \textit{s}, and idler \textit{i} tones. Here, $\tilde{k}_{j}$ denotes a general wave vector that is frequency and power dependent. Whereas the rule of energy conservation is inherently satisfied, satisfying the second condition presents greater difficulty due to the inherent curvature of the dispersion relation in a Josephson Junction (JJ) array and the self- and cross-Kerr effects (or equivalently, the self- and cross-phase modulations), which results in power-dependent vectors $\tilde{k}$. It is critical to emphasize that momentum conservation can significantly influence the amplification characteristics of a TWPA, resulting in gain variations ranging from 10 dB to 20 dB between a device lacking momentum matching considerations and one that is meticulously engineered for such purposes, despite the nonlinear properties being identical \citep{planat_photonic-crystal_2020,planat_resonant_2020}.

\begin{figure}[h]
\includegraphics[scale=0.6]{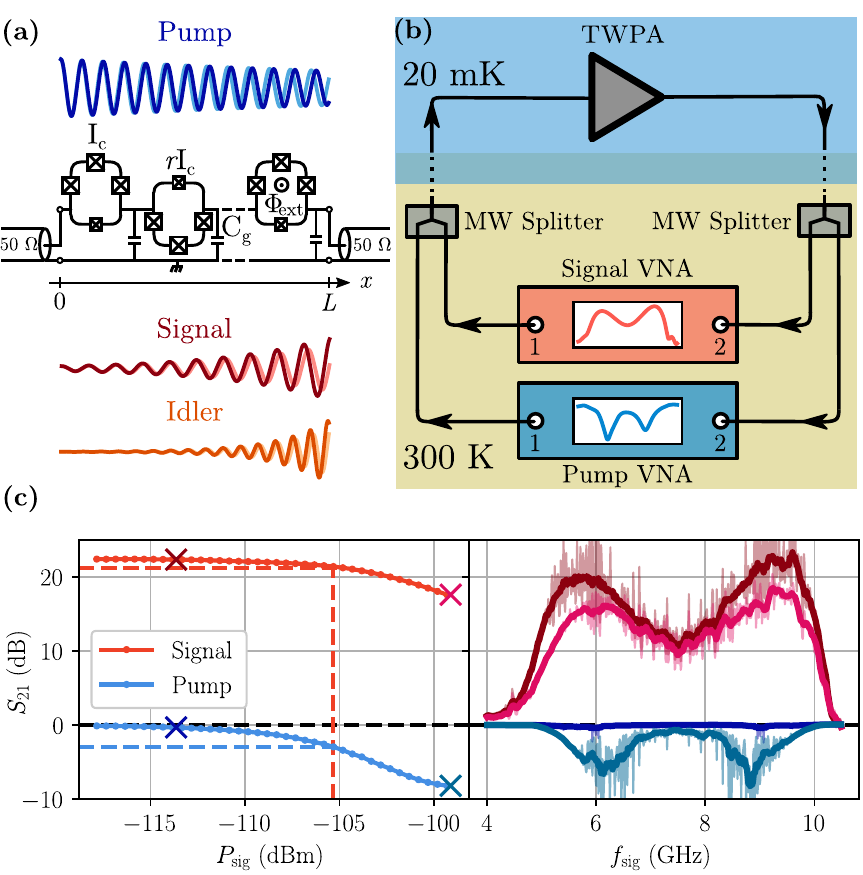}

\caption{\label{fig:Fig_1}\textbf{1-dB compression in a TWPA.} \textbf{(a)}
The device consists in a chain of SNAIL forming a $50\:\mathrm{\Omega}$
matched non-linear transmission line. The device is flux biased at
$\Phi_\mathrm{ext}=\Phi_{0}/2$. A strong pump tone (blue) at frequency $\omega_{p}$
provides the energy to amplify the signal tone (red) at frequency
$\omega_{s}$ and generate an idler tone (orange)
at frequency $\omega_{i}=2\omega_{p}-\omega_{s}$. The tones can experience a phase shift as they propagate due to self and cross-phase modulation
effects (highlighted by the light-color waves), in addition to being amplified
or attenuated. The amplitudes of all these effects are exaggerated for illustrative purposes. \textbf{(b)}
Simplified schematic of the room temperature experimental setup. The
microwave source usually used to generate the pump tone is replaced
by a second VNA (\textquoteleft pump VNA\textquoteright ). \textbf{(c)}
Examples of experimental curves obtained from the experimental setup
shown in (b). Signal gain and pump transmission (red and blue, respectively)
as a function of the input signal power $P_{\mathrm{sig}}$ (left
panel) for $f_{\mathrm{sig}}=$ 6 GHz and $f_{\mathrm{pump}}=$ 7.5
GHz, and versus signal frequency (right panel). The pump power is
approximately $-78.4$~dBm. The vertical dashed line shows $P_{\mathrm{1dB}}$,
and the blue horizontal dashed line highlights pump attenuation at $P_{\mathrm{sig}}=P_{1\mathrm{dB}}$.
The dark colors correspond to $P_{\mathrm{sig}}=-113.6$~dBm, and
the brighter colors to $P_{\mathrm{sig}}=-99.3$~dBm (see corresponding
colored crosses in the left panel). The light shades show the raw
acquired data while the thicker plain lines show the smoothed transmission
profiles.}
\end{figure}

The device under study here, whose schematic is shown in Fig. \ref{fig:Fig_1}, is a SNAIL-based TWPA \citep{frattini_3-wave_2017},
operated in the `reversed-Kerr' regime with an external
DC flux bias $\Phi_\mathrm{ext}=\Phi_{0}/2$ \citep{ranadive_kerr_2022}. Momentum matching is obtained by reversing the sign
of Kerr non-linearity, such that the power-induced modification of
the $\tilde{k}$ vectors compensates for the natural mismatch between
the zero-power or linear $k$ vectors. 

As schematically depicted in Fig. \ref{fig:Fig_1}(a),
energy is drawn from the pump to amplify the signal and generate the
idler tone. The different waves also inherit a phase shift due to self-
and cross-phase modulation effects as their amplitudes vary while
they propagate and interact. In this picture, these phenomena can limit the achievable gain of
the device for two reasons. First, as the amplitudes of the signal and idler waves increase, some pump power is \deleted{irreversibly} converted to $\omega_{s}$ and
$\omega_{i}$ and less energy is available for amplification, until
the gain saturates before the waves exit the TWPA (this effect is known
as \textit{pump depletion}). Second, as the wave amplitudes vary,
momentum mismatch, or phase-mismatch, due to Kerr effects can increase, and the gain can
also saturate inside the device because of decreasing interaction efficiency. This effect can be triggered by pump depletion.

In practice, it has already been shown that as long as the input signal power
is small with respect to the pump one, such effects are negligible and optimal performance of the TWPA is achieved \citep{yaakobi_parametric_2013,obrien_resonant_2014,planat_resonant_2020,ranadive_kerr_2022}.
However, when the input signal power is increased, the amplified signal
tone inside the device can become a significant fraction of the pump
one, and the effects mentioned above can significantly affect the
gain of the device, ultimately leading to compression, i.e. a significant
decrease of the signal gain compared to the linear one (obtained at low input signal power).

In order to experimentally study the causes of compression in a J-TWPA, we implemented the experimental setup shown in Fig. \ref{fig:Fig_1}(b).
It consists of a standard setup widely used to characterize the gain
of superconducting TWPA \citep{macklin_nearquantum-limited_2015,malnou_three-wave_2021,planat_photonic-crystal_2020,ranadive_kerr_2022,fadavi_roudsari_three-wave_2023},
where the pump and signal microwave tones are combined at room temperature
using microwave (MW) splitters. The signal tone is generated using a VNA, so that its frequency and power
can be varied while recording its transmission for different pump
parameters. The pump tone, usually generated
using a MW source, is generated with another VNA in our setup,
allowing to also record its complex transmission. The detailed experimental
setup is presented in the Appendix \ref{sec:Appendix-A:Exp_setup}.

The measurement procedure is the following: the frequency of the signal VNA
is swept while that of the pump is kept
constant. For each signal frequency, pump and signal transmissions are measured. This is repeated for several
signal powers. 

A typical data set obtained with this setup is shown in Fig. \ref{fig:Fig_1}(c).
The left panel shows the measured normalized transmission (defined below) $S_{21}$
of the signal tone, in red, and the one of the pump tone, in blue,
as a function of the signal power at the input of the TWPA. We will
refer to gain when $S_{21}$~\replaced{>}{<}~0~dB, and to depletion or attenuation
when $S_{21}$~<~0~dB. More precisely, we define the signal gain in
dB as $G=\bar{S}_{21}^{(s)}(\mathrm{pump\:ON})-\bar{S}_{21}^{(s)}(\mathrm{pump\:OFF})$,
where $\bar{S}_{21}^{(s)}$ is the raw measured transmission through
the whole setup with the signal VNA. Similarly, the pump transmission in dB
is defined as $S_{21}=\bar{S}_{21}^{(p)}(\mathrm{signal\:ON})-\bar{S}_{21}^{(p)}(\mathrm{signal\:OFF})$,
$\bar{S}_{21}^{(p)}$ is the raw measured transmission with the pump
VNA. The vertical dashed line shows the 1-dB compression
point defined as

\begin{equation}
P_{\mathrm{1dB}}=P_{\mathrm{sig}}(G_{\mathrm{lin}}^{\mathrm{dB}}-1\:\mathrm{dB})\label{eq:def_P_1db}
\end{equation}
where $G_{\mathrm{lin}}=G(P_{\mathrm{sig}}\rightarrow-\infty)$ is
the linear gain of the TWPA at low input signal power $P_{\mathrm{sig}}$.
One can clearly see that, as the signal gain drops, so does the pump
transmission. We attribute this attenuation to a non-negligible pump depletion. The right
panel shows gain profiles of the amplifier as a function of the signal
frequency for a low and higher $P_{\mathrm{sig}}$ in dark and bright
red, respectively. The dark and bright blue curves show the corresponding
pump transmission profiles as a function of the signal frequency and
same $P_{\mathrm{sig}}$. One clearly sees a significant attenuation
of several dB of the pump tone for signal frequencies close to the maxima of the corresponding signal gain.

In order to better understand these results, let us now present the
modeling of these data.

\section{Model of the gain compression}

\label{sec:Modeling_compression}The model we use is based
on the coupled-modes equations (CMEs) for the three modes envelopes' propagation. This model has already been
widely used to model the gain of different types of TWPA \citep{yaakobi_parametric_2013,obrien_resonant_2014,macklin_nearquantum-limited_2015,malnou_three-wave_2021,planat_resonant_2020,ranadive_kerr_2022}.
The more detailed derivation of the coupled mode equations for a
4WM SNAIL TWPA is recalled in the Appendix \ref{sec:Appendix-C:-Theoretical_modeling}.
We will here briefly recall the main steps up to the equations that we
solve numerically in this study. We will then give approximate analytical
formulae that were already derived in the literature \citep{kylemark_semi-analytic_2006,obrien_resonant_2014}.

As usual in this theory, we start from the classical equation of motion
for a generalized flux wave $\Phi(x,t)$ traveling in the device:

\begin {widetext}

\begin{equation}
\frac{\partial^{2}\Phi}{\partial x^{2}}+L(\Phi_\mathrm{ext})C_{J}\frac{\partial^{4}\Phi}{\partial x^{2}\partial t^{2}}-\frac{L(\Phi_\mathrm{ext})C_{g}}{a^{2}}\frac{\partial^{2}\Phi}{\partial t^{2}}-\frac{3a^{2}\tilde{\gamma}}{I_{c}^{2}L(\Phi_\mathrm{ext})^{2}\tilde{\alpha}^{3}}\frac{\partial^{2}\Phi}{\partial x^{2}}\left(\frac{\partial\Phi}{\partial x}\right)^{2}=0.\label{eq:eq_motion_SNAIL}
\end{equation}

\end {widetext}

Here, $L(\Phi_\mathrm{ext})$ is the flux-dependent inductance of the SNAIL
loop, $C_{J}$ is the equivalent Josephson capacitance of the SNAIL
loop, $C_{g}$ is the ground capacitance, $I_{c}$ the critical current
of the large JJ in the SNAIL loop, $a$ is the unit cell length in
meters, and $\tilde{\alpha}$, $\tilde{\gamma}$ are flux dependent
parameters coming from the Taylor expansion of the SNAIL current-phase
relation (see Appendix \ref{sec:Appendix-C:-Theoretical_modeling}
for further details). In order to obtain CMEs, we follow the method described in Appendix \ref{sec:Appendix-C:-Theoretical_modeling}.
However, unlike it is usually done to obtain simple analytical expressions
for the gain of the signal \citep{yaakobi_parametric_2013,obrien_resonant_2014,planat_resonant_2020},
we do not make any further approximation regarding the relative amplitudes
$A_{j}$ of each wave. We also consider the effect of distributed losses in the model. We finally obtain the following CMEs for the propagation of the three waves envelopes:

\begin {widetext}

\begin{equation}
\added{
\left\{\begin{aligned}
\frac{\partial A_s}{\partial x} &= i \alpha_{sp} A_p A_p^* A_s
                                + i \kappa_{si} A_p^2 A_i^* e^{i\Delta k_l x}
                                 + i \alpha_{ss} A_s^2 A_s^*
                                 + i \alpha_{si} A_i A_i^* A_s
                                 -k_s'' A_s\\
\frac{\partial A_i}{\partial x} &= i \alpha_{ip} A_p A_p^* A_i
                                 + i \kappa_{is} A_p^2 A_s^* e^{i\Delta k_l x}
                                 + i \alpha_{ii} A_i^2 A_i^*
                                 + i \alpha_{is} A_s A_s^* A_i
                                 -k_i''A_i \\
\frac{\partial A_p}{\partial x} &= i \alpha_{pp} A_p^2 A_p^*
                                 + i \alpha_{pi} A_i A_i^* A_p
                                 + i \alpha_{ps} A_s A_s^* A_p
                                 + i \kappa_{psi} A_s A_i A_p^* e^{-i\Delta k_l x}
                                 -k_p''A_p
\end{aligned}\right. .\label{eq:CPE_SNAIL}
}
\end{equation}

\end {widetext}

The coefficients $\alpha_{jj}$ are the self-Kerr coefficients, $\alpha_{jk}$
are the cross-Kerr coefficients, and the $\kappa$'s are the 4WM coefficients
ruling the energy exchange between the different waves. Their expressions
are given in Appendix \ref{sec:Appendix-C:-Theoretical_modeling}. \added{In the signal and idler equations, the two first terms are the usual ones kept within the stiff and strong pump approximations~\citep{obrien_resonant_2014}. The two following ones appear because we do not consider a strong pump ($A_s \sim A_i \lesssim A_p$). In the pump equation, only the first term is usually present within the approximations mentioned above~\citep{obrien_resonant_2014}, and the two following ones are also due to $A_s \sim A_i \lesssim A_p$. The fourth term is the one responsible for pump depletion and comes from the breakdown of the stiff pump approximation. It is essential to model our data.}
The \replaced{last}{first} term accounts for propagative losses via \added{the imaginary part of the wave-vector $k_{j}''=\tan(\delta)k_{j}/2$ (also corresponding to the real part of the propagation constant $\gamma = k'' + i k$)}, 
with $\tan(\delta)$ the loss tangent of TWPA,
and $\Delta k_{l}=2k_{p}-k_{s}-k_{i}$ is the linear phase mismatch.
These equations are formally equivalent to those already derived
for optical parametric amplifiers \citep{chen_four-wave_1989,cappellini_third-order_1991,kylemark_semi-analytic_2006},
or superconducting TWPA \citep{yaakobi_parametric_2013,obrien_resonant_2014}.
It should be noted that, with further approximations, analytical formulae
can be obtained for propagation $A_{j}(x)$ including pump depletion effects \citep{chen_four-wave_1989,obrien_resonant_2014}. \added{Similar expressions for three-wave mixing TWPA were also derived~\citep{zorin_josephson_2016}.}
However, we will not use them in this work as they do not account for losses, and rely on Jacobi elliptic functions, thus not showing explicit dependence with the powers, non-linear coefficients, or phase-mismatch.

\begin{figure*}
\centering

\includegraphics{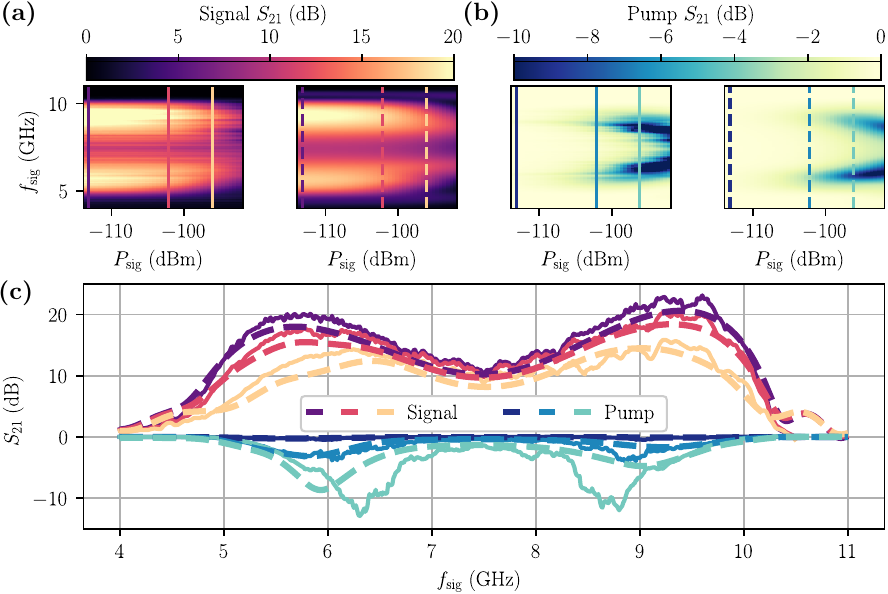}

\caption{\label{fig:Fig_2}\textbf{Comparison between experimental data
and theory. (a)}
Measured (left panel) and simulated (right panel) signal gain as a
function of the signal frequency and input signal power. \textbf{(b)}
Measured (left panel) and simulated (right panel) pump transmission
as a function of the signal frequency and input signal power. The
vertical lines correspond to the signal powers shown in (c). The simulations contain no fitting parameters and use the same input parameters for both (a) and (b). The axis
spans and color ranges are the same for experiment and theory in both (a) and (b). \textbf{(c)} Measured and simulated (plain and
dashed lines respectively) signal and pump transmissions (red and
blue shades respectively) as a function of the signal frequency for three input
signal powers: $P_{\mathrm{sig}} = -113.1$~dBm, $-102.1$~dBm, and $-96.1$~dBm (from dark to bright colors respectively). These curves correspond
to the vertical linecuts with respective colors in (a) and (b). The
experimental data shown in this figure are smoothed as in Fig. \ref{fig:Fig_1}(c). The pump power is $P_p = -78.4$~dBm.}

\end{figure*}

We then numerically solve these equations as a function of the position
$x$ in the TWPA, and we define the gain of the signal and the transmission of the pump
as 

\begin{equation}
\begin{array}{c}
\bar{S}_{21}^{(s)}=\left|A_{s}(x=L)/A_{s}(0)\right|^{2}\\
\bar{S}_{21}^{(p)}=\left|A_{p}(x=L)/A_{p}(0)\right|^{2}
\end{array}\label{eq:Gain_S21_def_thy}
\end{equation}
respectively. We also compute these transmissions without pump tone
($A_{p}(0)=0$), or signal tone ($A_{s}(0)=0$), to obtain the reference
for normalization of these 'calibrated' transmissions obtained numerically.
The difference between the simulations with signal (pump) tone and
the ones with $A_{p}(0)=0$ ($A_{s}(0)=0$) directly correspond to
the normalized experimental signal (pump) transmissions defined
in section \ref{sec:TWPA_experiment}, and shown in Fig. \ref{fig:Fig_1}(c).

In Fig. \ref{fig:Fig_2}, we show the results of these simulations,
obtained without any fitting parameters, compared to the experimental
data. All input parameters necessary to perform the simulations are the same for \ref{fig:Fig_2}(a) and (b), obtained from the linear characterization of the device presented in Appendix
\ref{sec:Appendix-B:linear_charac_device} and from the calibration
of the experimental setup presented in Appendix \ref{sec:Appendix-A:Exp_setup}.
We obtain a good quantitative agreement between the (smoothed) data
and the theory over a wide range of input signal powers, for both the
evolution of the gain profile (Fig. \ref{fig:Fig_2} (a)) and the
pump transmission profile (Fig. \ref{fig:Fig_2}(b)). We stress here
that the color ranges encoding the transmission values in the 2D plots
are the same for both data and theory. This agreement is also visible
in the linecuts shown in Fig. \ref{fig:Fig_2}(c). 

However, some more detailed features are not perfectly reproduced,
such as the exact bandwidth of the gain profile, or the frequencies
of the pump transmission minima. We attribute these discrepancies to
some simplifications of the model, where 
\replaced{perfect impedance matching with the environment}{perfectly power-independent $50\:\mathrm{\Omega}$ matching} is assumed, not accounting for the
stray geometric inductance of aluminum in the device, higher-order terms in the current-phase relation expansion, or position-dependent loss coefficients for example. 

In this figure, we also see two interesting features. First, the pump
transmission profile is asymmetric with respect to the pump frequency
(7.5 GHz). Second, the pump transmission is not minimum, where the
linear gain is maximum. This is observed both experimentally and theoretically.

Regarding asymmetry, it is mostly observed in the simulation results
as seen in Figs. \ref{fig:Fig_2}(a) and (b). Although it is seen
in the experimental evolution of the gain in Fig. \ref{fig:Fig_2}(a)
at higher signal powers, the noise and ripples in the pump transmission
measurements prevent one from concluding whether this behavior also shows
up clearly in the data. It is most likely due to
the frequency dependence of the losses, which are lower at smaller frequencies. This already gives
simulated asymmetric gain profiles for reversed Kerr TWPA without
introducing pump depletion in the model \citep{ranadive_kerr_2022}.
When energy exchange between the pump and the signal/idler is introduced,
the asymmetry in the gain profile is mapped onto the pump transmission profiles.

We will keep the discussion concerning the frequency difference between
the minimum of pump $S_{21}$ and the maximum of $G_{\mathrm{lin}}$
for the next sections, as it requires further discussion of compression
to be understood. We will now study in more detail the causes of
compression in TWPA.

\section{The 1-dB compression point}

\label{sec:Understanding_1dB}The model we presented earlier was utilized in initial theoretical studies to investigate the causes of compression in a 4WM-based TWPA \citep{yaakobi_parametric_2013,obrien_resonant_2014}.
In Ref. \citep{obrien_resonant_2014}, the main hypothesis to study
signal gain compression is that pump depletion is the only cause. The
authors show that, under this hypothesis and the absence of losses, the gain should evolve as 
\begin{equation}
G(P_{\mathrm{sig}})=\frac{G_{\mathrm{lin}}}{1+2G_{\mathrm{lin}}P_{\mathrm{sig}}/P_{\mathrm{p}}}\label{eq:G_vs_Psig_Obrien}
\end{equation}
where $G_{\mathrm{lin}}$ is the linear gain defined in section \ref{sec:TWPA_experiment},
$P_{\mathrm{p}}$ is the input pump power and $P_{\mathrm{sig}}$ is the input signal power. This formula should be valid
for any kind of 4WM-based TWPA \citep{cappellini_third-order_1991,kylemark_semi-analytic_2006,obrien_resonant_2014}.

\begin{figure}
\includegraphics{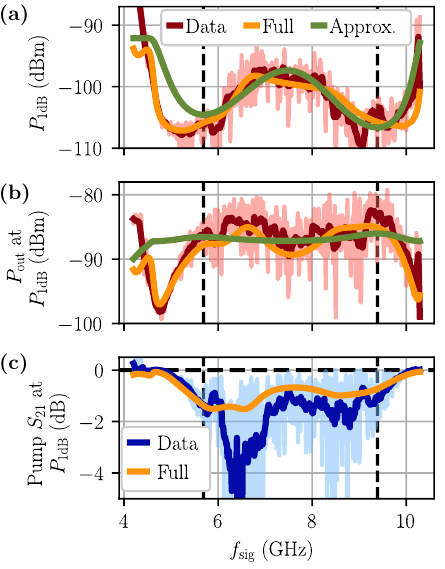}

\caption{\textbf{\label{fig:Fig_3}Frequency
dependence of the 1-dB compression point and pump transmission at
$\boldsymbol{P_{\mathrm{1dB}}}$. (a)} $P_{1\mathrm{dB}}$ vs signal
frequency. The red curve shows the experimentally measured $P_{1\mathrm{dB}}$
across the bandwidth of the TWPA. The orange curve shows the theoretical
$P_{1\mathrm{dB}}$ obtained from simulations with the model presented
in Appendix \ref{sec:Appendix-C:-Theoretical_modeling}. The green
curve shows $P_{1\mathrm{dB}}$ extracted from equation (\ref{eq:G_vs_Psig_Obrien})
where $G_{\mathrm{lin}}$ corresponds to the simulated gain with the
full model including losses at the lowest $P_{\mathrm{sig}}$. \textbf{(b)}
Output signal power at $P_{1\mathrm{dB}}$ as a function of the signal
frequency. The colors correspond to the same situations as in (a).
\textbf{(c)} The corresponding pump transmission profiles at $P_{\mathrm{sig}}=P_{1\mathrm{dB}}$
as a function of the signal frequency. The blue curve shows the experimental
data and the orange curve shows the simulated pump transmission from
the model of Appendix \ref{sec:Appendix-C:-Theoretical_modeling}.
In all the figures, the vertical dashed lines highlight the maxima
of the signal gain simulated using the full model at the lowest signal
power used experimentally. The brighter shaded lines show the raw
data while the darker curves correspond to the smoothed data. The pump power is $P_p = -78.4$~dBm.}
\end{figure}

In Fig. \ref{fig:Fig_3}(a) we show a comparison between the experimentally
obtained $P_{\mathrm{1dB}}$ (red curve), and the ones obtained from
the full simulations using the model of Appendix \ref{sec:Appendix-C:-Theoretical_modeling}, as well as the ones given by
Equation (\ref{eq:G_vs_Psig_Obrien}) in orange and green, respectively.
The linear gain $G_{\mathrm{lin}}$ used in equation (\ref{eq:G_vs_Psig_Obrien})
is the one simulated with the full model at the lowest input signal
power---hence accounting for losses. The method we used to extract
$P_{\mathrm{1dB}}$ from the data is given in Appendix \ref{sec:Appendix-D:data_analysis}. 

One can see that both theories give an overall good quantitative agreement 
with the data across the whole bandwidth of the device. However, the
two theoretical curves still display quite different features as a
function of signal frequency. They both match the data within the central part of the amplification
band (between the two vertical dashed lines), and claiming whether
one is more accurate than the other is not possible considering the
error bars of the measured $P_{\mathrm{1dB}}$ (represented
by a light-shaded red curve). They are, however, quite different in the
low-frequency part of the amplification band, where the full model
matches the data well, while the simplified model does not.

A striking feature in both experiment and full theory is that $P_{\mathrm{1dB}}$
is not the lowest at the frequencies where the linear gain is maximum
(highlighted by the two vertical dashed lines). However, it is the
case for the approximated formula, equation (\ref{eq:G_vs_Psig_Obrien}),
on either side of the pump frequency (7.5 GHz). This can be understood
by noticing that the main feature of equation (\ref{eq:G_vs_Psig_Obrien})
is that $P_{\mathrm{1dB}}$ only depends on the linear gain $G_{\mathrm{lin}}$,
as already noted in \citep{obrien_resonant_2014}.
Therefore, the higher the linear gain, the lower $P_{1\mathrm{dB}}$.
This feature is also reproduced in Fig. \ref{fig:Fig_3}(b), where
we show the signal output power $P_{\mathrm{out}}$ at the input signal
power $P_{\mathrm{sig}}=P_{\mathrm{1dB}}$---i.e. $P_{\mathrm{1dB}}+G_{\mathrm{lin}}^{\mathrm{dB}}-1$
dB. The consequence is an almost flat $P_{\mathrm{out}}$ at $P_{\mathrm{1dB}}$
as a function of the signal frequency when obtained from equation
(\ref{eq:G_vs_Psig_Obrien}) (green curve), which again matches only
the data (red curve) within the center of the amplification band.
Conversely, the output powers at $P_{1\mathrm{dB}}$ obtained from
the simulations (orange curve) match well the data across the whole
band and seem to reproduce slight variations also observed experimentally. 

In Fig. \ref{fig:Fig_3}(c) we show the measured (blue) and simulated
(orange) pump transmission at $P_{\mathrm{sig}}=P_{\mathrm{1dB}}$.
The agreement between data and theory is also good (notice the scale
difference with respect to Figs. \ref{fig:Fig_3}(a) and (b)), except
for a big dip observed experimentally between $f_{\mathrm{sig}}=6-7$
GHz, which is most likely an artifact of the smoothing of the large ripples in the raw transmission profile. However, one clearly observes pump attenuation, \textit{viz.}
depletion, of about 1 dB or more, when $P_{\mathrm{sig}}=P_{\mathrm{1dB}}$.
This is observed across the entire bandwidth of the TWPA, except
on the edges of the band. This shows that compression 
is linked to pump depletion in general. However, it cannot explain the detailed behavior of TWPA at the saturation point, as we will discuss hereafter. Still, an interesting conclusion from the analysis of Figs. \ref{fig:Fig_3} is that although the expression (\ref{eq:G_vs_Psig_Obrien})
does not reproduce all the features observed experimentally, it still
gives a good quantitative estimate of $P_{1\mathrm{dB}}$ for a
4WM TWPA, provided that $G_{\mathrm{lin}}$ accounts for losses. However, since Eq. (\ref{eq:G_vs_Psig_Obrien}) was derived without accounting for them, it could be refined to make losses appear explicitly in gain dependence with signal power.

The main reason why $P_{\mathrm{1dB}}$ obtained from Eq. (\ref{eq:G_vs_Psig_Obrien})
is too approximate, is actually stated in Ref. \citep{kylemark_semi-analytic_2006}
where this expression was originally derived for optical fiber amplifiers.
It contains two approximations: (i) low input signal power compared
to the total input power, and (ii) valid in the limit of total
power conversion between the pump and signal/idler. While the first
one is valid even in the regime close to 1-dB compression ($P_{\mathrm{1dB}}<-90\:\mathrm{dBm}\ll P_{\mathrm{p}}=-78.4\:\mathrm{dBm}$),
the validity of the second one requires a more detailed analysis of the physics of
this non-linear system.

In order to understand with analytical formulae the second approximation,  no frequency dependence non-linear
coefficients (the equivalent of our $\alpha$'s and $\kappa$'s in
Eqs. (\ref{eq:CPE_SNAIL})) has also to be assumed. Following a Lagrangian approach,
this allows one to define an optimal, input power-dependent, linear phase mismatch
$\Delta k_{l}^{\mathrm{(opt)}}(P_{\mathrm{sig}},\:P_{\mathrm{p}})$
at the input of the amplifier that yields total, i.e. asymptotic,
power conversion between the pump and signal/idler waves \citep{cappellini_third-order_1991}.
In the case of our Josephson TWPA, this frequency dependence cannot
be neglected. Therefore, the Lagrangian formulation becomes more complex and it is difficult to obtain an equivalent of $\Delta k_{l}^{\mathrm{(opt)}}(P_{\mathrm{sig}},\:P_{\mathrm{p}})$
without similar assumptions. However, this explains
why Eq. (\ref{eq:G_vs_Psig_Obrien}) does not reproduce $P_{\mathrm{1dB}}$
and $P_{\mathrm{out}}$ at $P_{\mathrm{1dB}}$ in Figs. \ref{fig:Fig_3}(a)
and (b): it assumes that all frequencies are somehow perfectly phase
matched with maximum power conversion efficiency, which is obviously
not the case here \citep{ranadive_kerr_2022}.

\begin{figure}
\centering

\includegraphics{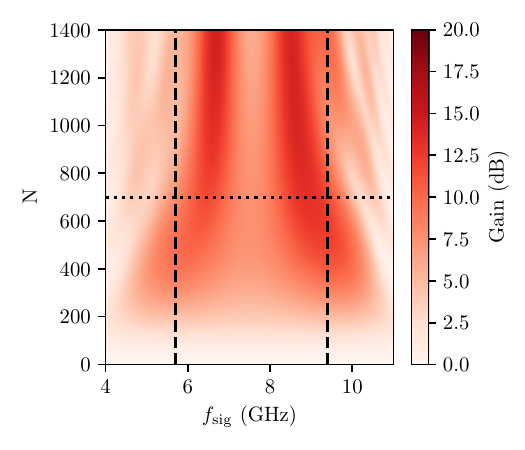}

\caption{\label{fig:Fig_4}\textbf{Stability of gain build-up at high signal power.} Simulated signal gain as a function of the position inside the TWPA
in unit cell number N and signal frequency. The input signal power is $P_{\mathrm{sig}}$ = -94.6 dBm. The horizontal dotted line indicates the actual
length of the device under study here (N = 700). The vertical dashed lines correspond to the two maxima of the simulated linear gain profile.}

\end{figure}

Let us now turn to the pump transmission profile at $P_{\mathrm{1dB}}$. Keeping in mind that not all frequencies are perfectly phase-matched in our device, even at low signal power, and that the total power-dependent phase-mismatch $\Delta \tilde{k}$ is mainly a function of pump power in J-TWPAs \citep{macklin_nearquantum-limited_2015}, an effect of pump depletion is to modify $\Delta \tilde{k}$ during propagation. Thus, on the edges of the band (outside the vertical
dashed lines), where no significant pump attenuation is observed,
$P_{\mathrm{1dB}}$ in Fig. \ref{fig:Fig_3}(a) is as low as for the frequencies where $G_{\mathrm{lin}}$ is maximum. Since edge frequencies are already initially largely phase mismatched
at low $P_{\mathrm{sig}}$ by definition, very small variations of
the power of any wave during propagation mismatches even more the amplification process,
yielding saturation of the gain before the output of the device. The effect of this `depletion-induced' matching modification is illustrated in Fig. \ref{fig:Fig_4}, where two regions can be identified. In the central part of the bandwidth, there is a stable frequency range where the gain increases monotonically with position inside the device. Conversely, the edges of the band are an instable region where the gain oscillates with position. In other words, for small $\Delta \tilde{k}$ the gain as a function of position is a monotonically increasing function (typically $\propto \cosh{(gx)}$ where $g$ is some coefficient and $x$ the position~\citep{obrien_resonant_2014}), but it becomes a periodic function of position if $| \Delta \tilde{k} |$ is too large yielding an imaginary $g$ \citep{obrien_resonant_2014}. In the specific case of a SNAIL-TWPA operated in the reversed-Kerr regime, depleting the pump tends to lower $\Delta \tilde{k}$, making the central frequencies better phase-matched, but mismatching the frequencies more detuned from the pump. The boundaries between the two regions are difficult to infer but are close to the maxima of the linear gain (vertical dashed lines in Fig. \ref{fig:Fig_4}), and result from the parameters of the device, pump power and the losses. This yields a complex interplay between lowering the energy available for amplification, and power-induced phase-mismatch with possible coherent, i.e. oscillatory, energy exchange between waves from which results gain compression. \added{In principle, the central region should also show a coherent behavior in an infinitely long TWPA~\citep{yaakobi_parametric_2013}.}

It should be noted that a TWPA with a single Josephson junction as unit cell might display a different behavior for all frequencies as in that case $\Delta \tilde{k}$ is always negative with our conventions, and lowering pump power would shift it towards zero. This also stands for a dispersion engineered TWPA like the one of Ref. \citep{macklin_nearquantum-limited_2015}. This highlights that the coherent nature of the amplification process inside a TWPA should be considered for accurate modeling of the gain at high input signal power~\citep{yaakobi_parametric_2013}.
\added{Finally, in the case of an ideal three-wave mixing amplifier, only pump depletion should cause gain compression because self-phase and cross-phase modulation effects are absent.}

\section{Conclusion and discussions}

\label{sec:Conclusion-and-discussions}In this article, we have presented
an experimental study of compression in a SNAIL traveling-wave parametric
amplifier operated in 4WM. We have implemented a novel experimental
setup allowing one to track the pump tone transmission as a function of
the signal tone power and frequency. 

The study reveals that pump depletion is observed at 1-dB compression and input signal powers beyond.

Using the classical framework of coupled-mode equations, we are able
to accurately model the gain and pump transmission profiles for various
input signal powers, without any fitting parameters. The comparison between the data and theory shows
that a simplified formula (Eq. \ref{eq:G_vs_Psig_Obrien}) already obtained previously \citep{obrien_resonant_2014} with the hypothesis that solely pump depletion causes compression---an irreversible energy conversion argument, gives the good order of magnitude to model
the 1-dB compression point of such a TWPA. However, one needs to simulate
the entire system beyond some of the approximations yielding Eq. (\ref{eq:G_vs_Psig_Obrien})
to capture more detailed features of the frequency dependence of $P_{\mathrm{1dB}}$ and the corresponding pump transmission profile. This treatment encompasses the effect of losses and power-induced variations of the phase-matching $\Delta \tilde{k}$ during propagation yielding oscillatory behavior of the modes envelopes. When the measured
signal gain and pump transmission decrease monotonically as a function
of input signal power, compression is mainly due to pump
depletion where significant linear gain is observed. Yet, pump-depletion induced $\Delta \tilde{k}$ causes compression for the frequencies being already largely phase-mismatched at low input signal power.

It is straightforward that the main limitation is the amount
of pump power that can be fed to the device. Therefore, in order to
increase the 1-dB compression point of 4WM-based J-TWPAs, Josephson junctions with larger
critical current should be used. Engineering of the TWPA (dispersion
relation, optimal length) should obviously be performed to compensate
for the resulting lower non-linearity of each unit cell.

Second, given that the bandwidth of TWPA is large and the gain usually
not constant, so is $P_{\mathrm{1dB}}$. Therefore, for applications
such as multiplexed qubit readout where the large input power comes
from several low-power signals at different frequencies, these tones
could be concentrated in frequency bands where $P_{\mathrm{1dB}}$
is higher, such as frequencies closer to the pump in the case
of the SNAIL-TWPA studied here. In this way, one stays far from the regime close to compression with significant generation of unwanted intermodulation products and sidebands \citep{remm_intermodulation_2023}. \added{Additionally, understanding how the variations in frequency of $P_{\mathrm{1dB}}$ affects the generation of third-order intermodulation products generation between far detuned signal tones across the band is an important tasks to further improve TWPAs for multiplexed qubit readout.} 

We believe that our study confirms existing theoretical predictions and will help designing TWPAs with larger 1-dB compression points, paving the way towards efficient setups for multiplexed qubit readout, a key requirement for scaling up quantum circuits.

\section*{Acknowledgments}

This project has received funding from the European Union's Horizon 2020 research and
innovation programme within the project AVaQus (grant agreement number 899561), from European Union's Horizon Europe 2021-2027 project TruePA (grant agreement number 101080152), and from the French ANR-22-PETQ-0003 grant under the `France 2030' plan.
B.F. acknowledges the QMIC project under the program DOS0195438/00.
We would like to acknowledge M. Esposito for her significant assistance with the sample fabrication.
The sample was fabricated at the clean room facility \textit{Nanofab} of Institut N\'eel in Grenoble. 
We thank the clean room staff and L. Cagnon for their assistance with the device fabrication.
We express our gratitude to, J. Jarreau, L. Del Rey, D. Dufeu, F. Balestro, and W. Wernsdorfer for their support with the experimental equipment. 
We are also grateful to the superconducting quantum circuits group members at Institut N\'eel for helpful discussions.
We thank Q. Ficheux and  R. Albert for their critical reading and constructive feedback on the manuscript.

G.L.G., G. B., and N.R. conceptualized the experiment. A.R. fabricated the device. G.L.G. and G.B. measured in the cryogenic
setup with the help of A.R., B.F. and G.C. L.P., E.B., E.E. and N.R. provided support with the
measurement setup. G.L.G.  performed data curation with the help of G.B., A.R., B.F., G.C. and N.R. G.L.G. performed the formal analysis. G.L.G., and N.R. drafted
the article with contributions from all the authors.

\added{The data presented in this article are openly available~\citep{data_zenodo_compression_resub_may_2025}.}
\appendix

\section{Experimental setup}

The device is mounted in a copper box and connected to
a $50\:\mathrm{\Omega}$ copper coplanar waveguide (CPW) using Aluminum
wire bonds. The box is anchored along with a coil for DC flux bias
to the mixing chamber of a dilution fridge and operated at a temperature
close to 20~mK. It is placed inside a $\mathrm{\mu}$-metal magnetic
shield. The amplifier is followed by two single-junction isolators,
and a High-Electron-Mobility-Transistor (HEMT) low-noise amplifier
anchored at the 4K stage of the fridge.

\begin{figure*}
\centering

\includegraphics{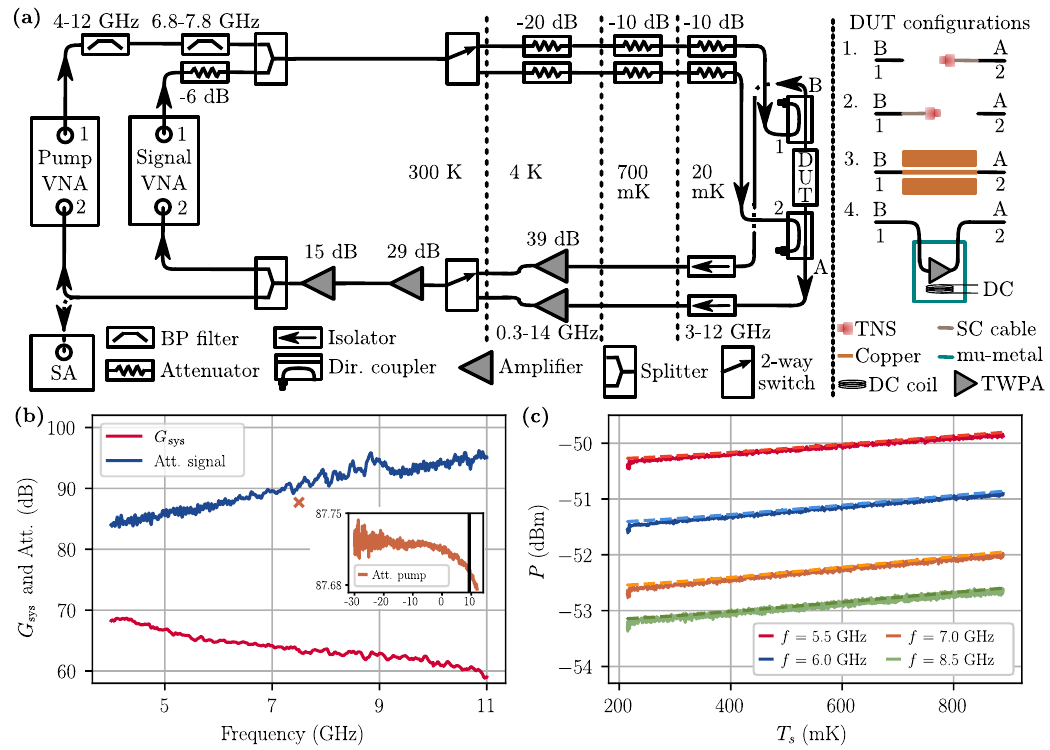}

\caption{\label{fig:Fig_A1}\textbf{Experimental setup and system calibration.
(a)} Full experimental setup. `DUT' (device under test) corresponds
to the different configurations depicted on the right and explained
in the main text. BP: band-pass, SA: spectrum analyzer, SC: superconducting,
TNS: thermal noise source (a cryogenic 50 $\Omega$ cap), VNA: vector
network analyzer. For simplicity, we represent only one isolator but
two are used on each line to provide $\sim$40 dB isolation. A \&
B label the outputs while 1 \& 2 label the inputs on the right panel. The directional couplers have 20 dB attenuation on the coupled port.
\textbf{(b)} System gain (red) as a function of frequency and the
resulting input line attenuation for the signal line (blue) obtained
as $S_{21}$ measured with configuration 3 in (a) minus $G_{\mathrm{sys}}$
(both in dB). The orange cross shows the attenuation of the pump line
at the pump frequency $f_{p}=7.5$~GHz. \textbf{Inset:} Input line
attenuation in dB for the pump line as a function of the pump VNA power
in dBm. The vertical line shows the VNA power used for all the measurements
presented in this article. \textbf{(c)} Examples of power emitted by
the TNS when varying the temperature of the stage it is anchored to
(plain lines), along with the fits (dashed lines) used to obtain $G_{\mathrm{sys}}$
in (a) from Eq. (\ref{eq:TNS_fit}).}

\end{figure*}

\label{sec:Appendix-A:Exp_setup}The complete experimental setup used
in this study is shown in Fig. \ref{fig:Fig_A1}(a). As compared to
usual setups used for characterizing TWPA, we use a second VNA at
room temperature for generating the pump tone and measuring its transmission
as discussed in the main text (section \ref{sec:TWPA_experiment}).
In addition, at the base temperature of the dilution fridge, the sample
is `sandwiched' between two directional couplers at input and output
in order to measure the full scattering matrix of the device. Reflection
measurements are not used in this study. 

In order to perform all the calibration needed for the study, we did
several cooldowns with different samples:
\begin{enumerate}
\item Thermal noise source at output A and an open cable at output B. This
allows calibrating the system gain of output line A and the base reflection
of output line B.
\item Open cable at output A and thermal noise source on output B. This
is the reciprocal of the first cooldown.
\item Dummy `PCB' sample in between the two directional couplers. It is
an on-chip 50 $\mathrm{\Omega}$ matched copper coplanar-waveguide
transmission line. The packaging used is the same as for the TWPA
(copper box, connectors and wire-bonding). It is used to calibrate
the transmission of the lines and provides a reference for the linear
characterization of the TWPA as discussed in section \ref{sec:Appendix-B:linear_charac_device}.
\item SNAIL TWPA in between the two directional couplers with its coil to
apply DC magnetic flux to the sample.
\end{enumerate}
In between all these cooldowns, nothing else than the different devices
facing the two directional couplers was modified.

In order to obtain the input line attenuation of our input line and
estimate accurately the powers at the input of the TWPA, we used a
thermal noise source to calibrate our output system gain. We then
subtract it to the full transmission across the fridge measured on
the PCB sample to obtain the input line attenuation including losses
of the packaging. The error made on this estimation is thus solely
the intrinsic losses of the PCB transmission line and on one side
of the packaging.

The thermal noise source is a cryogenic 50 $\mathrm{\Omega}$ cap,
and is anchored to a copper mount held by a thin copper wire providing
a thermal weak link to the mixing chamber plate. The RF output of
the cap is connected to the outputs of the fridge using a superconducting
micro-wave cable. In this way, it could thermalize down to 200 mK
during the cooldown, and when heated up to calibrate the gain, it does
not heat up in return the mixing chamber plate. A thermometer and a heater
are also anchored to the copper mount.

To perform the calibration, we measure the noise power emitted by
the source at different temperatures. We then fit the data for each
frequency using the model for the measured power
in watts at room temperature:

\begin{equation}
P(\omega)=\left(N_{\mathrm{source}}+N_{\mathrm{sys}}\right)G_{\mathrm{sys}}\Delta fk_{B}\label{eq:TNS_fit}
\end{equation}
where $\Delta f$ is the integration bandwidth, $N_{\mathrm{sys}}$ and $G_{\mathrm{sys}}$
are respectively the noise temperature (in K) and total gain of the
system. The noise temperature in K emitted by the thermal source ($N_{\mathrm{source}}$)
is modeled as

\begin{equation}
N_{\mathrm{source}}=\frac{\hbar\omega}{2k_{B}}\coth\left(\frac{\hbar\omega}{2k_{B}T_{s}}\right)\label{eq:Noise_temp_TNS}
\end{equation}
where $T_{s}$ is the source temperature. The result of the gain calibration
along with estimation of the input line attenuation is shown in Fig.
\ref{fig:Fig_A1}(b).

We show in Fig. \ref{fig:Fig_A1}(b) some examples of these fits,
as well as the resulting system gain calibration and estimated input-line
attenuation for both the signal line and the pump line (different
because of the extra attenuator used with the signal VNA at room temperature
as shown in Fig. \ref{fig:Fig_A1}(a)).

\section{Device fabrication and linear characterization}

\label{sec:Appendix-B:linear_charac_device}

\subsection*{Fabrication of the device}

The device under study here was fabricated in the clean room of the Institut N\'eel. The fabrication steps are identical to those of
the device presented in Ref. \citep{ranadive_kerr_2022}.
We here briefly recall the main steps. 

First, the Josephson junctions and Aluminum bridges constituting the
SNAIL array are patterned using electron-beam lithography and evaporated using a double-angle shadow evaporation Dolan bridge technique. Then,
a thin layer ($\sim 25$ nm) of Aluminum Oxyde is deposited on the
whole wafer using atomic-layer deposition. Finally, a thick layer of
Copper is deposited on top of the wafer, except for the bounding
pads, to form the ground capacitance of the array. 

\subsection*{Linear characterization}

In order to estimate all the parameters necessary to model the data
we measure the linear parameters of the device. To do so, we measure
the device's transmission as a function of the flux threading the
SNAIL loops. We then subtract the transmission measured on the
PCB sample, assumed to be lossless with respect to the device's attenuation.
The $S_{21}$ parameter's magnitude and phase are normalized
in order to obtain the losses as a function of frequency and the linear
dispersion relation, respectively:

\begin{equation}
\left|S_{21}(\Phi_\mathrm{ext})\right|=\left|S_{21}^{(\mathrm{DUT})}(\Phi_\mathrm{ext})\right| / \left|S_{21}^{(\mathrm{PCB})}\right|,\label{eq:S21_norm_lin_carac}
\end{equation}

\begin{equation}
\theta(\Phi_\mathrm{ext})=\theta^{(\mathrm{DUT})}(\Phi_\mathrm{ext})-\theta^{(\mathrm{PCB})}.\label{eq:Phase_norm_lin_carac}
\end{equation}

\begin{figure*}
\centering

\includegraphics{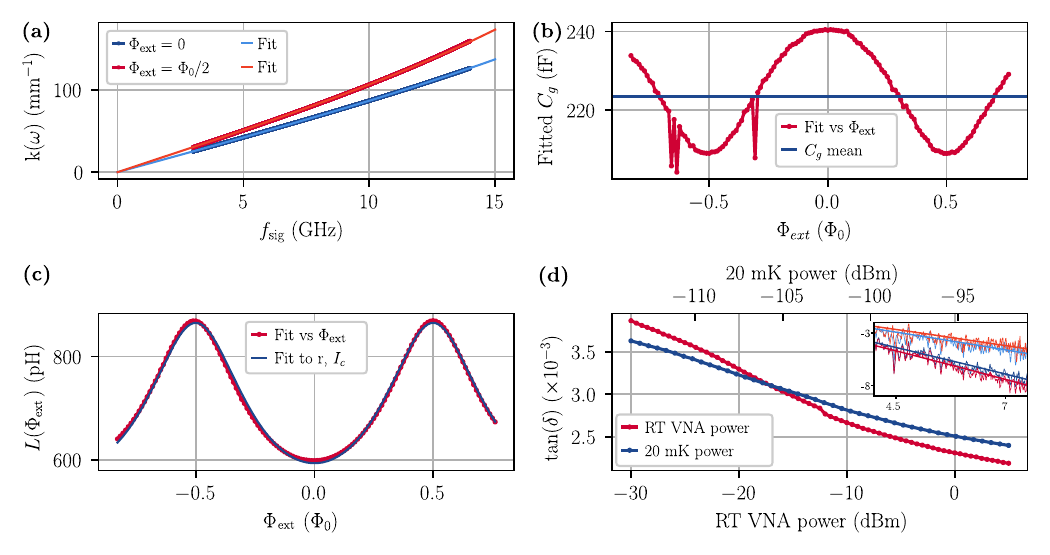}

\caption{\label{fig:Fig_A2}\textbf{Linear characterization of the device.
(a)} Dispersion relation of the device extracted from phase measurements
at two external magnetic fluxes along with the corresponding fits
using Eq. (\ref{eq:k_disp}) with only $L(\Phi_\mathrm{ext})$ as fitting
parameters. \textbf{(b)} Fitted ground capacitance $C_{g}$ versus
$\Phi_\mathrm{ext}$ with 2 fitting parameters in Eq. (\ref{eq:k_disp})
and the mean value used as a fixed parameters for the final fit of
$L(\Phi_\mathrm{ext})$ and the simulations. \textbf{(c)} Fitted values $L(\Phi_\mathrm{ext})$
versus $\Phi_\mathrm{ext}$ with only one fitting parameter in Eq. (\ref{eq:k_disp})
in red along with the fit to estimate $r$ and $I_{c}$ in blue from
Eq. (\ref{eq:flux_dep_L_SNAIL}). \textbf{(d)} Fitted $\tan(\delta)$
from $\left|S_{21}\right|$ versus $k(\omega)$ for different input
powers. Data in red correspond to fits of raw transmissions for different
RT VNA powers (bottom axis), while blue data correspond to fits of reconstructed
transmissions for similar input power at each frequency
(upper axis) as explained in Appendix \ref{sec:Appendix-D:data_analysis}.
\textbf{Inset:} Raw $\left|S_{21}\right|$ data in dB versus frequency in
both cases (same color code) for the lowest power (dark colors) and
the largest power (brighter colors) of the main figure along with
the transmission calculated from the respective fitted $\tan(\delta)$.}

\end{figure*}

We first fit $\theta(\Phi_\mathrm{ext})$ with a general function for the
dispersion relation:

\begin{equation}
\theta(\Phi_\mathrm{ext})=\theta_{0}(\Phi_\mathrm{ext})+\frac{N\omega\sqrt{L(\Phi_\mathrm{ext})C_{g}}}{\sqrt{1-L(\Phi_\mathrm{ext})C_{J}\omega^{2}}}\label{eq:theta_fit_gen}
\end{equation}
where $\theta_{0}(\Phi_\mathrm{ext})$, $L(\Phi_\mathrm{ext})$, $C_{g}$ and $C_{J}$
are fitting parameters. As an outcome of this fit we only retain the
parameter $\theta_{0}(\Phi_\mathrm{ext})$ and use it to rescale the absolute
phase of the first measured frequency of the VNA traces. The absolute
wave vectors $k(\Phi_\mathrm{ext})$ are then defined as 
\begin{equation}
k(\Phi_\mathrm{ext})=\frac{\theta(\Phi_\mathrm{ext})+\theta_{0}(\Phi_\mathrm{ext})}{l},\label{eq:k_from_phase}
\end{equation}
where $l=Na$ is the length of the device with $N$ the number of
SNAIL and $a$ the length of one unit cell. Then, we fit the dispersion
relation $k(\omega)$ for each flux value using the standard formula

\begin{equation}
k(\omega)=\frac{\omega\sqrt{L(\Phi_\mathrm{ext})C_{g}}}{a\sqrt{1-L(\Phi_\mathrm{ext})C_{J}\omega^{2}}}\label{eq:k_disp}
\end{equation}
with only $L(\Phi_\mathrm{ext})$ and $C_{g}$ as fitting parameters. $C_{J}=31$~
fF is known from fabrication and corresponds to the equivalent Josephson
capacitance of the SNAIL loop. As an outcome of this fit, a flux-dependent
$C_{g}$ is obtained as shown in Fig. \ref{fig:Fig_A2}(b), which
is not physical as the dependence with flux of $k(\omega)$ is expected
to be only due to the variation of the inductance $L(\Phi_\mathrm{ext})$.
We therefore fix $C_{g}$ to the mean value obtained from this fit,
and fit again the dispersion relation with equation (\ref{eq:k_disp})
and $L(\Phi_\mathrm{ext})$ as the only fitting parameter. The result is
shown in Fig. \ref{fig:Fig_A2}(a).

In order to estimate the ratio $r$ and the critical current of the
large junctions $I_{c}$, we then fit the dependence of the inductance
with the external flux with the expected variation of the linear inductance
of the SNAIL loop from the current-phase relation expansion:

\begin{equation}
L(\Phi_\mathrm{ext})=\frac{\varphi_{0}}{\tilde{\alpha}(\Phi_\mathrm{ext})I_{c}}\label{eq:flux_dep_L_SNAIL}
\end{equation}
where $\tilde{\alpha}(\Phi_\mathrm{ext})$ is defined in equation (\ref{eq:alpha_tilde})
and $\varphi_{0}=\hbar/(2e)$ the reduced flux quantum. The fit is
shown in Fig. \ref{fig:Fig_A2}(c) and we obtain $r=0.062$ and $I_{c}=1.4\:\mathrm{\mu A}$.

The last parameter needed for the simulations is the loss tangent
$\tan(\delta)$. \deleted{Before describing the procedure to extract it, let
us make a few comments. The main loss process we expect for our device
is dielectric losses located in the dielectric forming the top ground
(see above for fabrication details and Ref. \citep{planat_fabrication_2019}).
In superconducting circuits operated in the microwave regime, such
losses are usually attributed to the excitation of two-level systems
(TLS) which brings power-dependent losses \citep{planat_fabrication_2019}.
In our device, this power dependence is also observed, and the typical
evolution of $\left|S_{21}\right|$ versus power at a given frequency
is very similar to the one of the internal quality factor of a resonator
versus the average number of photons inside. Since we extensively
study the properties of our device over a broad range of powers (more
than 20~\replaced{dB}{dBm}), we therefore need to account for the power dependence
of the loss tangent to accurately model our data. In addition, we
need it for the signal and idler, but also for the pump whose power
is much larger than the one of the signal.}

In order to obtain the most accurate estimation for the pump at the right power and for the
signal at all the powers presented in this article, we extract $\tan(\delta)$
on two different kinds of data with the same procedure. \added{The reasons for using power-dependent loss tangents in the simulations are discussed in Appendix \ref{sec:Appendix-E:tan_delta_gain}.} Let us first
present the procedure. 

The attenuation in a transmission line is usually modeled as follows
in linear units \citep{planat_fabrication_2019,pozar_microwave_2012}:

\begin{equation}
A(x,\omega)=e^{\alpha_{d}(\omega)x}\label{eq:Attenuation}
\end{equation}
with 
\begin{equation}
\alpha_{d}(\omega)=\frac{k(\omega)\tan(\delta)}{2}.\label{eq:alpha_d}
\end{equation}
Therefore, $\tan(\delta)$ can be obtained by fitting the transmission
versus $k(\omega)$. This definition of the attenuation coefficient
is in Np/m. Thus, from the fit of $\left|S_{21}(\omega)\right|$
in dB versus $k(\omega)\times L$ (or equivalently the measured absolute
phase) with a linear relation $S_{21}^{\mathrm{dB}}=k(\omega)L\alpha_{\mathrm{fit}}$ we
obtain the loss tangent as defined above as:

\begin{equation}
\tan(\delta)=\alpha_{\mathrm{fit}}\times2\ln(10)/20\label{eq:tan_delta}
\end{equation}
where $\alpha_{\mathrm{fit}}$ is the fitted slope in dB/rad.

The results of these fits for transmission traces as a function of
the RT VNA power are shown in Fig. \ref{fig:Fig_A2}(d). To estimate
$\tan(\delta)$ at the pump power used, we take the value obtained
at the largest RT VNA power used, the closest to the pump power\footnote{We could not measure directly transmission traces with the pump VNA
to estimate more precisely $\tan(\delta)$ at the pump power used
in the experiments.}. 

To estimate the value of $\tan(\delta)$ at the different powers used
for the signal, we fit composite $\left|S_{21}\right|$ versus phase
data, where each $\left|S_{21}\right|$ trace is reconstructed so
that for each frequency the input power at the device is the same
as explained in Appendix \ref{sec:Appendix-D:data_analysis}. In both
cases however, the phase data used for these fits are the same for
all the powers, coming from the dataset used to fit the dispersion
relation shown in Fig. \ref{fig:Fig_A2}(a) at $\Phi_\mathrm{ext}=\Phi_{0}/2$.
This choice was made to avoid possible (although negligible in our
data) phase shift due to self-Kerr effect that could modify $k(\omega)$
and add a correction to $\tan(\delta)$ while we want to obtain the
linear loss tangent. 

\section{Theoretical modeling based on coupled-mode equations}

\label{sec:Appendix-C:-Theoretical_modeling}The model we use to compare
to our data is based on the widely used coupled-mode equations (CME)
for simulating the gain of the TWPA. The equation of motion (\ref{eq:eq_motion_SNAIL})
is obtained by applying Kirchoff's current law at each node of the
lumped-element model of the SNAIL chain. We use the following current
phase relation for the SNAIL:

\begin{equation}
I=rI_{c}\sin(\Phi/\varphi_{0})+I_{c}\sin\left(\frac{\Phi-\Phi_\mathrm{ext}}{3\varphi_{0}}\right)\label{eq:curr_phase_SNAIL}
\end{equation}
where $\Phi=\varphi_{0}\phi$ is the generalized flux for $\phi$
the superconducting phase difference across the SNAIL. It is further
expanded around a flux $\Phi^{*}$ such that $I(\Phi^{*})=0$ as

\begin{multline}
I(\Phi^{*}+\delta\Phi)\approx\frac{I_{c}}{\varphi_{0}}\left[r\cos\left(\frac{\Phi^{*}}{\varphi_{0}}\right)+\frac{1}{3}\cos\left(\frac{\Phi^{*}-\Phi_\mathrm{ext}}{3\varphi_{0}}\right)\right]\delta\Phi\\
-\frac{I_{c}}{6\varphi_{0}^{3}}\left[r\cos\left(\frac{\Phi^{*}}{\varphi_{0}}\right)+\frac{1}{27}\cos\left(\frac{\Phi^{*}-\Phi_\mathrm{ext}}{3\varphi_{0}}\right)\right]\delta\Phi^{3}+O\left(\delta\Phi^{5}\right)\label{eq:taylor_cur_phase_rel_SNAIL}
\end{multline}
where we already dropped the terms independent of $\delta\Phi$ and
proportional to $\delta\Phi^{2}$, anticipating that we will work
at $\Phi_\mathrm{ext}=\Phi_{0}/2$ where the coefficients are zero \citep{frattini_optimizing_2018,ranadive_nonlinear_2023}.
By defining the coefficients $\tilde{\alpha}$ and $\tilde{\gamma}$
as 

\begin{equation}
\tilde{\alpha}=r\cos\left(\frac{\Phi^{*}}{\varphi_{0}}\right)+\frac{1}{3}\cos\left(\frac{\Phi^{*}-\Phi_\mathrm{ext}}{3\varphi_{0}}\right)\label{eq:alpha_tilde}
\end{equation}

\begin{equation}
\tilde{\gamma}=\frac{1}{6}\left[r\cos\left(\frac{\Phi^{*}}{\varphi_{0}}\right)+\frac{1}{27}\cos\left(\frac{\Phi^{*}-\Phi_\mathrm{ext}}{3\varphi_{0}}\right)\right]\label{eq:gamma_tilde}
\end{equation}
and $L(\Phi_\mathrm{ext})=L_{J}/\tilde{\alpha}=\varphi_{0}/(\tilde{\alpha}I_{c})$
this yields the equation of motion (\ref{eq:eq_motion_SNAIL})
in the continuous limit. In practice, $\tilde{\alpha}$ and $\tilde{\gamma}$
are obtained by solving equation (\ref{eq:curr_phase_SNAIL}) at the
$\Phi_\mathrm{ext}$ of interest. 

Then, we use the following Ansatz to solve equation (\ref{eq:eq_motion_SNAIL}):

\begin{equation}
\Phi=\frac{1}{2}\underset{j\:\in\:\{p,s,i\}}{\sum}\left(A_{j}(x)e^{i(k_{j}x-\omega_{j}t)}+A_{j}^{*}(x)e^{-i(k_{j}x-\omega_{j}t)}\right),\label{eq:Ansatz_phi}
\end{equation}
where $A_{j}$ is the slowly varying amplitude in units of flux (Wb) of
the mode $j$ along $x$; $p,\:s,\:i$ respectively denote the pump,
signal and idler modes with $k_{j}$ and $\omega_{j}$ their associated
wave-vectors and angular frequencies. To solve the equation of motion, we perform the following approximations:
\begin{itemize}
\item $|\partial^{2}A_{j}/\partial x^{2}|\ll|k_{j}\partial A_{j}/\partial x|$
in the linear terms,
\item $|\partial^{2}A_{j}/\partial x^{2}/k_{j}|,|\partial A_{j}/\partial x|\ll|k_{j}A_{j}|$
in the non-linear term,
\item rotating-wave approximation by forcing the energy conservation $2\omega_{p}=\omega_{s}+\omega_{i}$
and retaining only the time-independent terms,
\end{itemize}
to obtain the CME (\ref{eq:CPE_SNAIL}) for the spatial variation
of the three modes envelopes and their complex conjugates. In these
equations, the non-linear coefficients are defined as

\begin{equation}
\alpha_{jm}=\frac{6\tilde{\gamma}}{\tilde{\alpha}^{3}}\dfrac{a^{4}k_{m}^{2}k_{j}^{3}}{8C_{g}I_{c}^{2}L(\Phi_\mathrm{ext})^{3}\omega_{j}^{2}}\;j\neq m,\label{eq:alpha_cross}
\end{equation}

\begin{equation}
\alpha_{jj}=\frac{6\tilde{\gamma}}{\tilde{\alpha}^{3}}\dfrac{a^{4}k_{j}^{5}}{16C_{g}I_{c}^{2}L(\Phi_\mathrm{ext})^{3}\omega_{j}^{2}},\label{eq:alpha_self}
\end{equation}

\begin{equation}
\kappa_{jm}=\frac{6\tilde{\gamma}}{\tilde{\alpha}^{3}}\dfrac{a^{4}k_{s}k_{i}k_{p}^{2}(2k_{p}-k_{m})}{16C_{g}I_{c}^{2}L(\Phi_\mathrm{ext})^{3}\omega_{j}^{2}},\label{eq:kappa_sig_idl}
\end{equation}

\begin{equation}
\kappa_{psi}=\frac{6\tilde{\gamma}}{\tilde{\alpha}^{3}}\dfrac{a^{4}k_{s}k_{i}k_{p}^{2}(k_{s}+k_{i}-k_{p})}{8C_{g}I_{c}^{2}L(\Phi_\mathrm{ext})^{3}\omega_{p}^{2}},\label{eq:kappa_pump}
\end{equation}
and the linear phase-mismatch $\Delta k_{l}=2k_{p}-k_{s}-k_{i}$.
One can notice that these coefficients are formally equivalent those derived for a single-JJ chain in Ref. \citep{obrien_resonant_2014}
up to replacing the Josephson inductance $L_{J}$ by $L(\Phi_\mathrm{ext})$,
and introducing the flux dependent normalization factor $\xi=6\tilde{\gamma}/\tilde{\alpha}^{3}$.
We introduce propagation losses for each wave phenomenologically by
adding a damping term in the RHS of the CME (\ref{eq:CPE_SNAIL})---which
amounts to neglecting the effect of losses on the non-linear coupling
of the waves and only consider them for the overall propagation, where
the imaginary component of the wave-vectors are defined as

\begin{equation}
k_{j}''(\omega)=\tan(\delta)k_{j}(\omega)/2.\label{eq:kj_loss}
\end{equation}

We then decompose each equation into differential equations onto their
real and imaginary parts and solve them numerically versus position
$x$ inside the device, with all the parameters in SI units. When
comparing to data, we take the last point of the simulation in $x=L$.
The initial flux amplitudes are defined as

\begin{equation}
A_{s,p}(0)=\sqrt{P_{\mathrm{sig,p}}Z_{0}}/\omega_{s,p},\:A_{i}(0)=0\label{eq:Init_cond_flux}
\end{equation}
with $Z_{0}=\sqrt{L(\Phi_\mathrm{ext})/C_{g}}$ the characteristic impedance
of the TWPA.

We finally summarize in Tab. \ref{tab:Sim_param} the linear parameters
used for the simulations.

\bigskip{}

\begin{table}
\centering

\begin{tabular}{|c|c|c|}

\hline 
\textbf{Parameter} & \textbf{Value} & \textbf{Unit}\tabularnewline
\hline 
Number of SNAIL $N$ & 700 & ---\tabularnewline
\hline 
Unit cell length $a$ & 8.7 & $\mathrm{\mu m}$\tabularnewline
\hline 
Critical current ratio $r$ & 0.062 & ---\tabularnewline
\hline 
Large junctions critical current $I_{c}$ & 1.4 & $\mathrm{\mu A}$\tabularnewline
\hline 
SNAIL inductance $L(\Phi_\mathrm{ext})$  & 869.6 & pH\tabularnewline
\hline 
SNAIL capacitance $C_{J}$ & 31 & fF\tabularnewline
\hline 
Ground capacitance $C_{g}$ & 223.5 & fF\tabularnewline
\hline 
Pump power $P_{\mathrm{p}}$ & $-78.4$ & dBm\tabularnewline
\hline 
Pump loss tangent $\tan(\delta)$ at $P_{\mathrm{p}}$ & 2.19$\times10^{-3}$ & ---\tabularnewline
\hline 
Signal loss tangent $\tan(\delta)$ & see Fig. \ref{fig:Fig_A2}(d) & ---\tabularnewline
\hline 
External magnetic flux $\Phi_\mathrm{ext}$ & $\Phi_{0}/2$ & Wb\tabularnewline
\hline 
\end{tabular}

\caption{\label{tab:Sim_param}Linear parameters of the TWPA used for the simulations.
All the flux-dependent parameters are estimated at $\Phi_\mathrm{ext}=\Phi_{0}/2$. }
\end{table}

\vfill{}

\section{Data analysis methods}

\label{sec:Appendix-D:data_analysis}In order to obtain the data as
presented in Figs. \ref{fig:Fig_2} and \ref{fig:Fig_3}, several
processing steps were performed from the raw data measured with both
VNA. We here detail the different steps.

\subsection*{Single input power gain profiles reconstruction}

Because of the frequency dependence of the input line attenuation
of the fridge at microwave frequencies, a measured gain or pump transmission
VNA trace for a given input signal power at room temperature (RT)
actually corresponds to different input signal powers referred at the
input of the TWPA. In order to ease the data analysis and compare
equivalently all the frequencies in the TWPA bandwidth, we chose to
reconstruct the gain and pump transmission profiles for a given input
signal power referred at the input of the TWPA. 

\begin{figure*}
\centering

\includegraphics{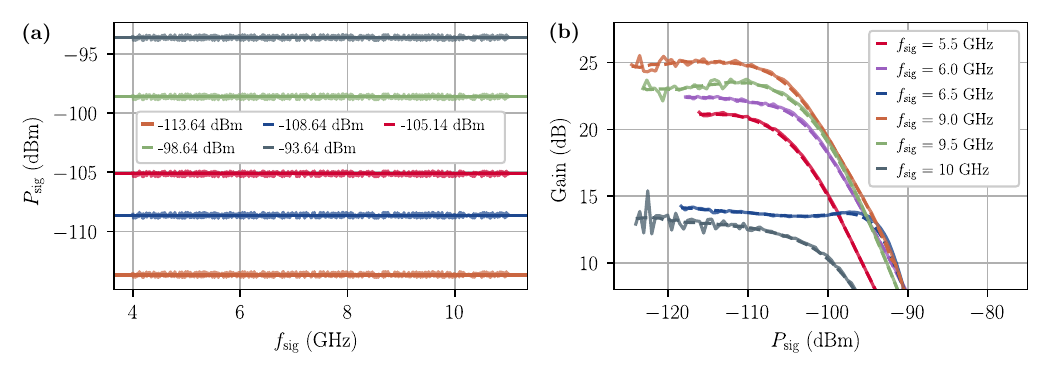}

\caption{\label{fig:Fig_A3}\textbf{Data processing scattering. (a)} Examples
of input signal powers across the whole measurement bandwidth and
their averages (horizontal plain lines) used to define the input signal
powers in this work. \textbf{(b)} Examples of raw (plain lines)
and smoothed (dashed lines) signal gain $G$ as a function of the
input signal power $P_{\mathrm{sig}}$ for different frequencies.}

\end{figure*}

To do so, we first rescale the VNA output powers at the input of the
device thanks to the system gain calibration presented in Appendix
\ref{sec:Appendix-A:Exp_setup}. We therefore obtain for each RT power
an array of powers $P_{\mathrm{TWPA}}(\omega)$ at the input of the
device containing the frequency dependence---or equivalently a matrix $\mathcal{P}(P_{\mathrm{TWPA}},\omega)$. Since the attenuation
globally increases with frequency (see Fig. \ref{fig:Fig_A1}(b)),
we then select the maximum power of the $P_{\mathrm{TWPA}}(\omega)$
array corresponding to the smallest RT power, defining the minimum
power at the input of the device $P_{\mathrm{sig}}^{\mathrm{(min)}}$
possibly shared by all the frequencies. In the same fashion, we select
the minimum power of the $P_{\mathrm{TWPA}}(\omega)$ array corresponding
to the highest RT power. This defines the maximum power $P_{\mathrm{sig}}^{\mathrm{(max)}}$
shared by all frequencies. Then, for each frequency, we truncate the
corresponding power array to retain all the powers at the input of
the device in the range $\left(P_{\mathrm{sig}}^{\mathrm{(min)}},P_{\mathrm{sig}}^{\mathrm{(max)}}\right)$---i.e.
truncating $\mathcal{P}(P_{\mathrm{TWPA}},\omega)$ along the power direction. Doing so, we basically
discard all the powers at the input of the TWPA that are not approximately
available at all frequencies (there is scattering because of the non
exactly linear input line attenuation). We then define each single
input signal power at the input of the device $P_{\mathrm{sig}}$
as the average power along the frequency axis of the truncated matrix $\mathcal{P}(P_{\mathrm{TWPA}},\omega)$ restricted to
the range $\left(P_{\mathrm{sig}}^{\mathrm{(min)}},P_{\mathrm{sig}}^{\mathrm{(max)}}\right)$.
It results an array of powers $P_{\mathrm{sig}}$ where transmission
data were measured for each of them, or a power close-by. In Fig. \ref{fig:Fig_A3}(a),
we show some iso-power cuts of the truncated matrix $\mathcal{P}(P_{\mathrm{TWPA}},\omega)$ as a function of frequency
before averaging as well as the averaged power (horizontal plain lines).
One can see that the power scattering is very small, not exceeding
$\pm0.5$ dBm.

We then reconstruct the complex transmission profiles by taking, for
each frequency, the raw VNA data at the power at the input of the
device the closest to the averaged one $P_{\mathrm{sig}}$. 

\subsection*{Extraction of $P_{\mathrm{1dB}}$}

As it can be seen in Fig. \ref{fig:Fig_1}(c) of the main text, the
raw gain data are quite noisy and exhibit ripples. Therefore, the
resulting curves of gain as a function of input signal power can be
quite noisy, mostly at low powers, hindering a consistent and efficient
detection of the 1-dB compression point for all the studied frequencies.
In order to extract it with more accuracy and automatically for each
frequency, we apply a moving average filter on the G vs $P_{\mathrm{sig}}$
curves. This smooths the low-power noise. Then, the 1-dB compression
is defined by equation (\ref{eq:def_P_1db}). Fig. \ref{fig:Fig_A3}(b)
shows a comparison between some typical raw $G$ vs $P_{\mathrm{sig}}$
curves (plain lines) and the smoothed result (dashed lines). We also
apply the same treatment to the quantities studied at $P_{\mathrm{1dB}}$
(pump $S_{21}$ at $P_{\mathrm{1dB}}$ shown in Fig. \ref{fig:Fig_3}(c)
for instance).

\subsection*{Averaged profiles versus signal frequency}

In order to compare more consistently the data to the theory and smooth
the scattering due to the ripples of the device, we also average the
different quantities studied as a function of signal frequency. To
this end, the measured transmissions when $f_{\mathrm{sig}}=f_{\mathrm{pump}}$
are removed---this avoids a staircase-like step around $f_{\mathrm{pump}}$
because of the high transmission measured at this frequency, and an 11-points moving-average filter is applied on the data. The result on the various quantities
studied can be seen in Figs. \ref{fig:Fig_1}(c) and \ref{fig:Fig_3}
by comparing the shaded curves and the plain lines.

\section{Effect of $\tan (\delta)$ on gain simulations}

\label{sec:Appendix-E:tan_delta_gain}
\begin{figure}

\includegraphics[scale=0.75]{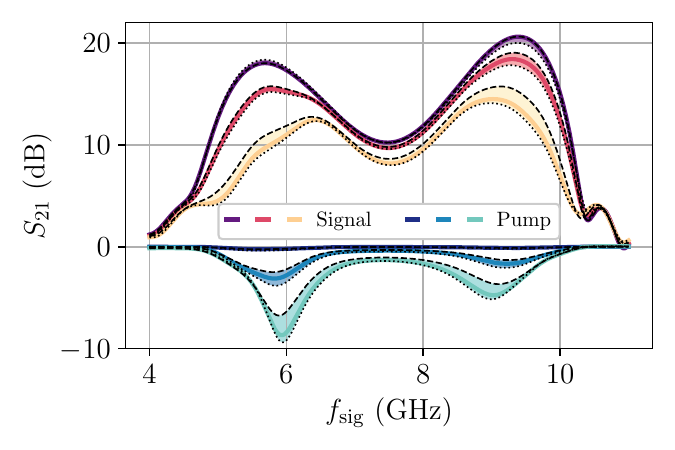}

\caption{\label{fig:Fig_A4}\textbf{Effect of the choice of $\tan(\delta)$ for modeling the TWPA gain.} Simulated signal gain and pump attenuation as a function of pump frequency for the same input signal powers as in Fig.~\ref{fig:Fig_2}(d). The plain lines correspond to the same simulations as Fig.~\ref{fig:Fig_2}(d), i.e. with power-dependent signal and idler $\tan(\delta)$. The shaded area correspond to the two limiting case of the lowest and largest values of $\tan(\delta)$ shown in Fig.~\ref{fig:Fig_A2}(d) for 20~mK power. The dotted and dashed lines highlight the boundaries of the two limits respectively.}

\end{figure}

\added{We discuss here the choice of using a power-dependent $\tan(\delta)$ at signal and idler frequencies for the simulations.
The main loss process we expect for our device
is dielectric losses located in the dielectric forming the top ground
(see Appendix~\ref{sec:Appendix-B:linear_charac_device} for fabrication details and Ref. \citep{planat_fabrication_2019}).
In superconducting circuits operated in the microwave regime, such
losses are usually attributed to the excitation of two-level systems
(TLS) which causes power-dependent losses \citep{planat_fabrication_2019}.
In our device, this power dependence is also observed, and the typical
evolution of $\left|S_{21}\right|$ versus power at a given frequency
is very similar to the one of the internal quality factor of a resonator
versus the average number of photons inside. Since we extensively
study the properties of our device over a broad range of powers (more
than 20~dB), we therefore need to account for the power dependence
of the loss tangent to accurately model our data. In addition, we
need it for the signal and idler, but also for the pump whose power
is much larger than the one of the signal. The procedure to extract $\tan(\delta)$ as a function of the power of the various tones is described in Appendix~\ref{sec:Appendix-B:linear_charac_device}.
However, it is known that a TLS can be saturated by applying a strong tone next to its resonant frequency. Therefore, it is natural to wonder about the frequency span over which the pump tone can saturate TLSs inside the TWPA to accurately model the losses at signal and idler frequencies.
According to the arguments for saturating TLSs using a strong pump tone presented in Ref.~\citep{capelle_probing_2020}, we estimate that the strong pump saturates TLSs over a frequency span of 100 to 200 MHz (assuming TLS dephasing rate of 1 MHz, $10^4$ photons for a traveling time of $\sim$10 ns, and a saturation number of photons $n_s \sim 20$). Therefore, we possibly over-estimate $\tan (\delta)$ for signal and idler only within a narrow frequency band around pump frequency. Thus, we consider that a power-dependent $\tan(\delta)$ for signal and idler gives a more accurate representation of the physics inside the device, meaning that the pump only saturates TLSs for frequencies nearby. We have checked that the choice of $\tan(\delta)$  for signal and idler within the range of Fig.~\ref{fig:Fig_A2}(d) has anyway a negligible impact on the simulation results and does not change the estimations of $P_{\mathrm{1dB}}$ and related quantities. The result is shown in Fig.~\ref{fig:Fig_A4} for the signal gain and pump attenuation. However, over-estimating $\tan(\delta)$ for the pump has a significant impact on the simulated gain profile, yielding significantly lower values.}

\bibliographystyle{apsrev4-2}
\bibliography{Bibliography}

\end{document}